\documentclass[journal]{IEEEtran}

\usepackage{graphicx}
\usepackage{amsmath}
\usepackage{amssymb}
\usepackage{gensymb}
\usepackage{amsthm}
\usepackage{placeins}
\usepackage{subcaption}
\usepackage{caption}
\usepackage{xcolor}
\usepackage{hyperref}
\usepackage{booktabs}
\usepackage{float}
\usepackage[ruled,vlined ]{algorithm2e}

\usepackage{bm}
\usepackage{multicol}
\usepackage{booktabs} 
\usepackage{xspace}
\usepackage{cite}

\usepackage{mwe}

\usepackage[ruled]{algorithm2e} 

\usepackage{xcolor}

\begin{document}

\title{Assessing Road Traffic Safety During COVID-19: Inequality, Irregularity, and Severity}  

\author{Lei Lin$^{1}$\thanks{${}^1$University of Rochester (E-mail: lei.lin@rechester.edu)}, Feng Shi$^{2}$\thanks{${}^2$University of North Carolina at Chapel Hill (E-mail: billanson10@gmail.com)}, Weizi Li$^{3}$\thanks{${}^3$University of Memphis (E-mail: wli@memphis.edu)}}

\maketitle

\begin{abstract} 
COVID-19 is affecting every social sector significantly, including human mobility and subsequently road traffic safety. In this study, we analyze the impact of the pandemic on traffic accidents using two cities, namely Los Angeles and New York City in the U.S., as examples. Specifically, we have analyzed traffic accidents associated with various demographic groups, how traffic accidents are distributed in time and space, and the severity level of traffic accidents that both involve and do not involve other transportation modes (e.g., pedestrians and motorists). We have made the following observations: 1) the pandemic has disproportionately affected certain age groups, races, and genders; 2) the ``hotspots'' of traffic accidents have been shifted in both time and space compared to time periods that are prior to the pandemic, demonstrating irregularity; and 3) the number of non-fatal accident cases has decreased but the number of severe and fatal cases of traffic accidents remains the same under the pandemic. 

\end{abstract}


\IEEEpeerreviewmaketitle

\section{Introduction}
\label{sec:intro}

The novel coronavirus 2019 (COVID-19) has undoubtedly impacted all aspects of our society~\cite{bonaccorsi2020economic,wang2020spatial}, with human mobility taking a big drop due to either voluntary isolation or governmental stay-at-home orders. A direct consequence of the reduced human mobility is reduced traffic, and hence a seemingly straightforward question follows: How is road safety affected by COVID-19? This simple question turns out to have conflicting answers. 

On one hand, the number of traffic accidents is positively correlated with the amount of traffic flows~\cite{shilling2020special}. Therefore, reduced mobility should result in a decreased number of accidents, which is confirmed by a few
recent studies. On the other hand, researchers and authorities both find that the new traffic pattern and the pandemic can lead to frequent speeding, careless driving, and even ``revenge driving,'' hence worsening road safety.

Here we take a systematic and data-driven approach to assessing road traffic safety during the COVID-19 pandemic. Road traffic safety is one of the most concerning topics in modern society. The total value of societal damage from motor vehicle crashes is estimated at \$871 billion dollars annually in the United States~\cite{crash-cost}. COVID-19 provides a natural experiment that greatly reduces traffic flows, and understanding how traffic accidents change accordingly will help us design more  effective road safety interventions. Furthermore, similar to COVID-19 cases which are disproportionately distributed among different demographic groups~\cite{poor}, the traffic accidents during COVID-19 are likely to show variations in various demographic factors, different times of day, and multiple severity levels. Knowing these variations also sheds a light on the heterogeneous impacts of COVID-19 and more generally, the transportation inequity analysis.


Previous studies have mainly focused on descriptive analysis using traffic accident data and used dates of government orders as the mobility change-point~\cite{oguzoglu2020covid, barnes2020effect,brodeur2020effects}. We, instead, use mathematical models and statistical tests analyzing the impact of the pandemic on road safety based on systematically detected mobility change-point in two cities: Los Angeles and New York City. 
We compare accidents before and after the mobility change-point using various demographic factors. We analyze the shift of accident ``hotspots'' in time and space. We also study the accident severity both involving and not involving other transportation modes. We extract three studying periods, namely 15-day, 30-day, and 60-day, before and after the mobility change-point to show the evolution of traffic accidents in terms of both absolute counts and fractions grouped by the aforementioned factors. In summary, we have made the following observations:

\begin{itemize}
    \item Although geographically distant, the detected mobility change-point of both Los Angeles and New York city, using Google Community Mobility Reports~\cite{Google}, is March 15, 2020.
    \item The pandemic is likely to affect different demographic groups unequally. We have analyzed the traffic accidents associated with age, race, and gender, both in terms of the counts and fractions under the pandemic. In terms of the age, all age groups except ``70--79,'' ``80--89,'' and ``90--99'' have significant reductions in accident counts. The groups ``20-29'' and ``30-39'' have the largest reductions. Regarding the fractions, only the groups ``10--19,'' ``20--29,'' and ``90--99'' in certain studying periods show significant changes. The largest reduction in fraction is seen in the youngest group ``10--19.'' 
    \item In terms of race, all groups have seen a reduction in accident counts. The ``Asian'' group has the lowest reduction but also the lowest daily accident counts, while the ``Hispanic'' group has the largest reduction but also the largest daily accident counts. Analyzing the fraction, the ``White'' group is the only group to have a reduction in all three studying periods.
    \item In terms of gender, we have three groups: ``Male'', ``Female'', and ``Other''. The accident counts of the ``Male'' and ``Female'' groups are reduced, while the ``Other'' remains the same. ``Female'' is the only group that has reduced accident fractions under the pandemic.
    \item Traffic accident counts have been reduced over most daytime and evening hours (i.e., between 6AM and 10PM) within 60 days after the mobility change-point. The fraction analysis shows that the ``peak hours'' of traffic accidents have shifted from morning (around 8AM) and afternoon (around 5PM) rush hours to late-evening hours (after 7PM). 
    \item In addition to time irregularity, we also find spatial irregularity as the traffic accident hotspots have shifted from their previous locations to different locations in both cities under the pandemic. 
    \item In terms of severity, for accidents that either involve or do not involve pedestrians or motorists, the overall counts are lower after the mobility change-point. However, surprisingly, the fatality counts remain the same, while the number of ``no injury'' and ``injury'' cases are reduced except for accidents that involve pedestrians. The fraction analysis shows that within all traffic accidents the ``no injury'' cases are proportionally higher than ``injury'' cases, but the difference becomes smaller or negligible 60 days after the mobility change-point.  
\end{itemize}


\section{Related Work}
\label{sec:relate}
We first review studies that analyze spatial-temporal characteristics of traffic accidents. Following that, we discuss recent work that assesses the impact of COVID-19 on road traffic safety. 

\subsection{Spatial and Temporal Traffic Accident Analysis}
The spatial-temporal analysis of traffic accidents aims to understand how, where, and when traffic accidents occur so that preventative actions can be executed to improve road traffic safety. From the spatial perspective, a group of studies has focused on identifying accident hotspots, i.e., locations where accidents occur frequently~\cite{xie2008kernel,kang2018spatiotemporal}, using various methods. For example, K-Means clustering algorithm is implemented to discover hotspots of pedestrian-involved crashes in Honolulu, Hawaii~\cite{kim2007using}. Community detection algorithm is adopted to reveal spatial-temporal features of accident hotspots  by first segmenting and then analyzing traffic accident data in heterogeneous groups~\cite{lin2014data}. Kernel Density Estimation (KDE), which can be used to estimate a spatial distribution of traffic accidents, has also been used for hotspot detection and visualization~\cite{anderson2009kernel,romano2017visualizing}. Compared to hard clustering algorithms such as K-Means, KDE can provide the likelihood of an event, thus modeling the uncertainty of an accident occurring at a location~\cite{anderson2009kernel}. 

From the temporal perspective, many studies have evaluated the effectiveness of various traffic safety interventions~\cite{robson2001guide,wahl2010red,missoni2012alcohol,green2016traffic, haghpanahan2019evaluation}. For example, Wahl et al. evaluate the efficacy of installing red light cameras (RLC) at intersections for traffic management. They find that RLC do not effectively prevent traffic accidents because of red-light running~\cite{wahl2010red}. Gree et al. demonstrate the effectiveness of a policy designed to reduce the number and severity of traffic accidents at London, which is to impose charges for traveling by car to the central city during peak hours~\cite{green2016traffic}. 

\subsection{COVID-19's Impact on Road Traffic Safety}
Due to the lockdown measures to contain and reduce the spread of COVID-19, domestic travel in the U.S. dropped 71\% in mid April compared to early March, 2020~\cite{trb2020}. Although the number of accidents is in general positively correlated with the amount of traffic flows~\cite{shilling2020special}, the number of fatalities, surprisingly, is observed to experience an increase at some states during this special time~\cite{vingilis2020coronavirus}. 
In particular, a report from the National Safety Council shows that the traffic accident fatality rate increased by 14\% in March 2020 compared to the same period in 2019~\cite{NSC2020}. 

Besides descriptively analyzing traffic accident rates~\cite{trb2020,shilling2020special,NSC2020}, other studies have mathematically modeled the effect of lockdown as a safety intervention~\cite{oguzoglu2020covid, barnes2020effect,brodeur2020effects}. To provide some examples, Barnes et al. adopt regression discontinuity design and find that, in Louisiana, U.S., while the overall number of accidents is reduced, more accidents are involving individuals from age 25 to 64, male, and nonwhite drivers~\cite{barnes2020effect}. Brodeur et al. use difference-in-differences to evaluate the impact of Stay-at-Home orders on traffic collisions for five states in the U.S. and find about 50\% reduction in traffic accidents after the implementation of the orders~\cite{brodeur2020effects}. 

The aforementioned studies all assume that people will adjust their mobility after governments issue guidelines and orders. However, depending on the study region and spatial granularity (e.g., state-level or nationwide), the mobility change could have an early response~\cite{villas2020we}. Currently, there lacks a systematic approach to detect the mobility change-point in existing studies and ignoring this aspect may lead to erroneous analysis of the pandemic's impact. In addition, none of the previously mentioned studies has explored the spatial distribution of accident hotspots during COVID-19, which knowledge could not only assist in designing more effective road safety interventions but also shed a light on transportation inequity analysis.


To overcome the limitations of the existing studies, we develop a change-point detection algorithm to capture when people change their mobility patterns in the pandemic. We also analyze the spatial distribution shift of traffic accidents, before and after the change-point date. Rigorous statistical tests are conducted in all our analyses. 


\begin{figure*}[ht]
        \centering
        \begin{subfigure}[b]{0.45\textwidth}
            \centering
            \includegraphics[width=\textwidth]{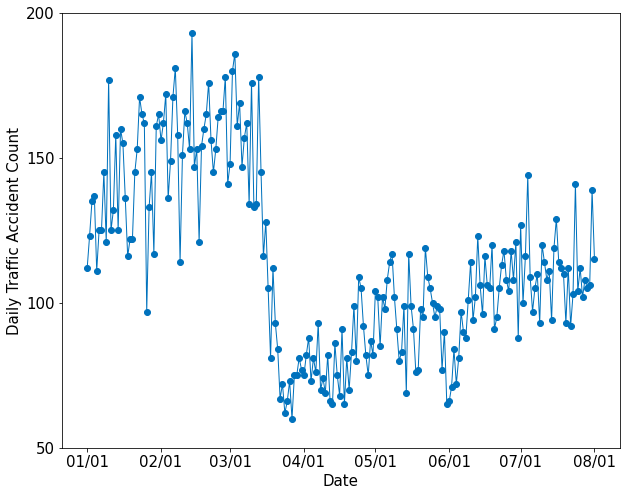}
            \caption[]{}%
            \label{fig:acc_la}
        \end{subfigure}\hspace*{0.9em}
        \begin{subfigure}[b]{0.45\textwidth}  
            \centering 
            \includegraphics[width=\textwidth]{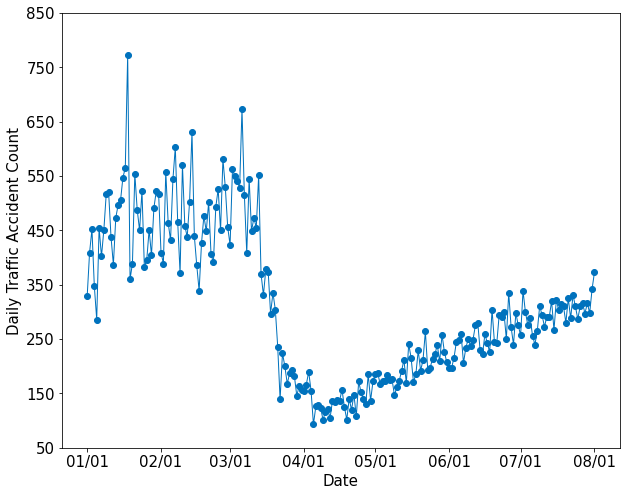}
            \caption[]{}%
            \label{fig:acc_nyc}
        \end{subfigure}
        \caption{Daily traffic accident count of (a) Los Angeles  and (b) New York City, between January 1 and August 1, 2020. Both cities have observed a significant deduction in traffic accidents in March. } 
        \label{fig:accident}
\end{figure*}
\section{Methods}
\label{sec:methods}
In this section, we introduce the main methods adopted in our analysis. 

\subsection{{Change-point Detection}} 
Consider a non-stationary time series $m=\{m_t\}_{t=1}^T$, which may have abrupt changes at $K$ unknown time steps $1<t_1<t_2<\cdots<t_K<T$. The goal of a change-point detection algorithm is to find these unknown time steps via solving the following optimization program:
\begin{equation}\label{eqn:opt}
    \min_{\tau} V(\tau) + \beta K,
\end{equation}
where $\beta$ is the weighting factor; $\tau = \{t_1,t_2,\cdots,t_K\}$ represents the segmentation of the time series. Both $\tau$ and $K$ are unknown and will be identified by the change-point detection algorithm. $V(\tau)$ is defined as:
\begin{equation}
    V(\tau) = \sum_{k=0}^{K}c(m_{t_k}..m_{t_{k+1}}), 
\end{equation}
where we additionally set $t_0 = 1$ and $t_{K+1} = T$; $c(\cdot)$ is the cost function that measures the similarity of the elements in the time series segment $m_{t_k}..m_{t_{k+1}} = \{m_t\}_{t_k}^{t_{k+1}}$. In this study, we choose $c(\cdot)$ to be the radial basis function~\cite{truong2020selective} and use the default value of $\beta$ as set in the ruptures Python package~\cite{truong2020selective}. 


\subsection{Difference-in-differences Analysis}
In order to statistically test the effect of lockdown on traffic accidents and the heterogeneity across different demographic groups, we perform difference-in-differences (DID) regression analyses. The data under study consist of daily accident counts of various demographic groups. We consider 15 days, 30 days, and 60 days before and after the detected mobility change-point as studying periods. 
In order to control seasonality patterns of traffic accidents, we also include data over the same periods in 2019 as the baseline.

Next, we quantify the relationship between daily accident counts and lockdown for different demographic groups using the following regression model:
\begin{equation}
    y \sim year*lockdown*x
\end{equation}
where $y$ denotes daily accident count; \textit{year} indicates either 2019 or 2020; \textit{lockdown} indicates whether a day is before or after the mobility change-point; and $x$ is a demographic variable (e.g., age, gender, race). Using gender as an example, $x$ can be either \textit{Male}, \textit{Female}, or \textit{Other} in this study. All three independent variables are categorical. The formula $year*lockdown*x$ considers not only individual variables but also their two-way and three-way interactions, i.e., $year + lockdown + x + year \times lockdown + \cdots + year\times lockdown\times x$. Following the convention of categorical variable analysis, the independent variables are converted into dummy variables, and the reference category is set to $year=2019$, $lockdown=No$, and $x=Male$. 

The results of the analysis are comprised of the coefficient of each three-way interaction term, which captures the difference in accident counts before and after the mobility change-point for each demographic group, subtracting the seasonality difference in 2019. 
We also report the confidence intervals of the coefficients for statistical significance. In general, a positive (negative) coefficient suggests an increase (decrease) of accidents with the value indicates the magnitude of change on average. In particular, if the confidence interval of a coefficient does not cover 0 or the corresponding $p$-value is less than or equal to 0.05, this coefficient is considered statistically significant; otherwise, insignificant.  

As the number of accidents is found to be correlated with the amount of traffic flows in general and decreasing after the mobility change-point, we further investigate whether the number of accidents changes disproportionally across various demographic groups. Specifically, 
the dependent variable is chosen to be either the daily accident count per group, or the fraction of daily accidents per group among all groups. We also report the resulting coefficients along with their confidence intervals.

\subsection{Kernel Density Estimation}
To study whether the locations of accidents have shifted as a result of changes in traffic patterns, we construct spatial distribution of the accidents by performing kernel density estimation on the accidents' locations. 

Taking the 30-day analysis as an example, we first split the data into four periods: 30 days before the mobility change-point in 2020, 30 days after the mobility change-point in 2020, and the corresponding periods in 2019. The two periods in 2019 serve as baselines to account for any seasonal shift of accidents' locations. Next, in each period, we fit the bivariate normal kernel to the (longitude, latitude) pairs of the accident data to pursue the estimation.

We provide both qualitative results through visualizing the resulting distributions and quantitative results in terms of statistical comparisons of the distributions. In particular, for the quantitative results, we conduct a global two-sample test on the accidents' locations with respect to the integrated squared error (ISE) between two density functions $f_1$ and $f_2$~\cite{duong2012closed}: 
\begin{equation}\label{eqn:ISE}
ISE = \int (f_1(x) - f_2(x))^2dx,
\end{equation}
where the null hypothesis is $H_0: f_1=f_2$. 
\section{Datasets}
\label{sec:data}

We use multiple datasets in this study. For detecting the mobility change-point, we use Google Community Mobility Reports~\cite{Google}. The dataset contains daily percentage changes of traffic with various purposes, including retail, recreation, grocery, pharmacy, and parks, at different regions. We use the combined percentages of all traffic as the metric to study the mobility trend of a region. 

Regarding traffic accidents, we use two datasets: one from Los Angeles~\cite{LATraffic} and the other from New York City~\cite{NYCTraffic}. Fig.~\ref{fig:accident} shows daily traffic accident counts of both cities between January 1 and August 1, 2020. We can see that both cities experience a reduction of traffic accidents in March 2020. Additionally, the dataset of Los Angeles records the time and location of each accident as well as demographic attributes such as age, severity, race, and gender. This enables us to study traffic accidents in different demographic groups. The studying periods are chosen to be 15 days, 30 days, and 60 days before and after the detected mobility change-point, i.e., March 15, 2020 (see next section for details).  
The dataset of New York City contains the time, location, and severity level of each accident. The severity is associated with two transportation modes, namely pedestrian and motorist, at three levels: no injury, injury, and fatality. Having this information, we can analyze the impact of COVID-19 on traffic accident severity. The same studying periods as before are adopted for analysis.



\section{Results}

\begin{figure*}
        \centering
        \begin{subfigure}[b]{0.5\textwidth}
            \centering
            \includegraphics[width=\textwidth]{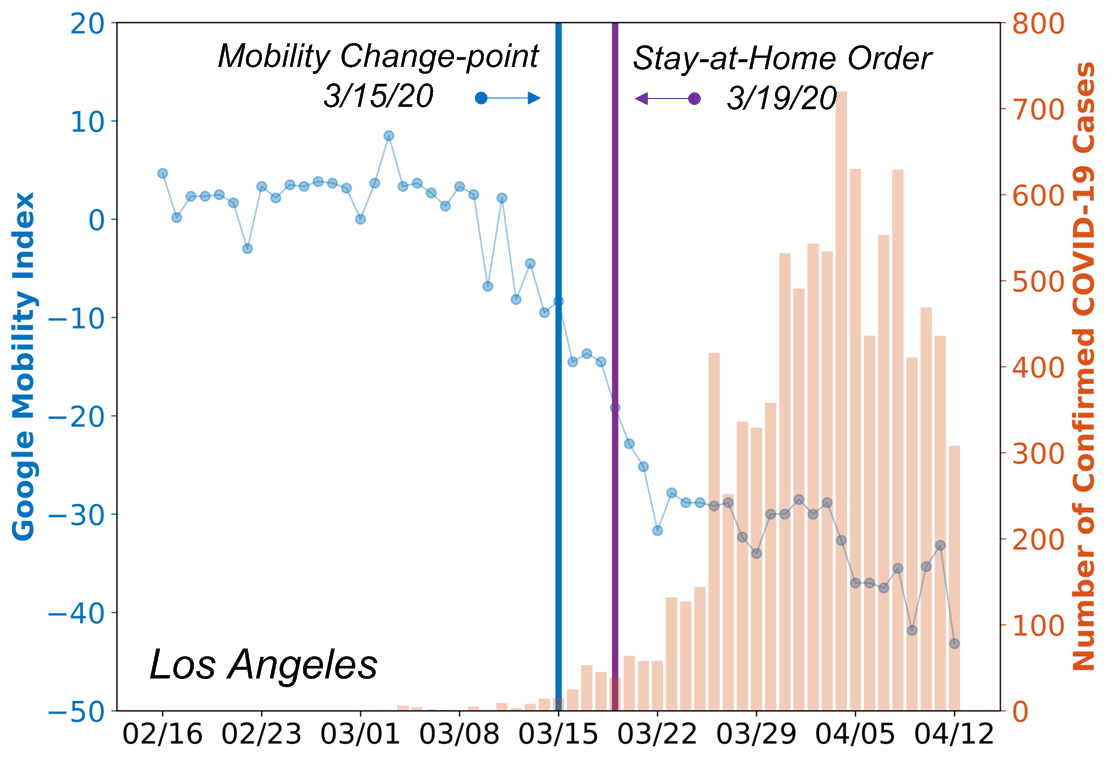}
            \caption[]{}%
            \label{fig:cp_la}
        \end{subfigure}\hspace*{0.9em}
        \begin{subfigure}[b]{0.5\textwidth}  
            \centering 
            \includegraphics[width=\textwidth]{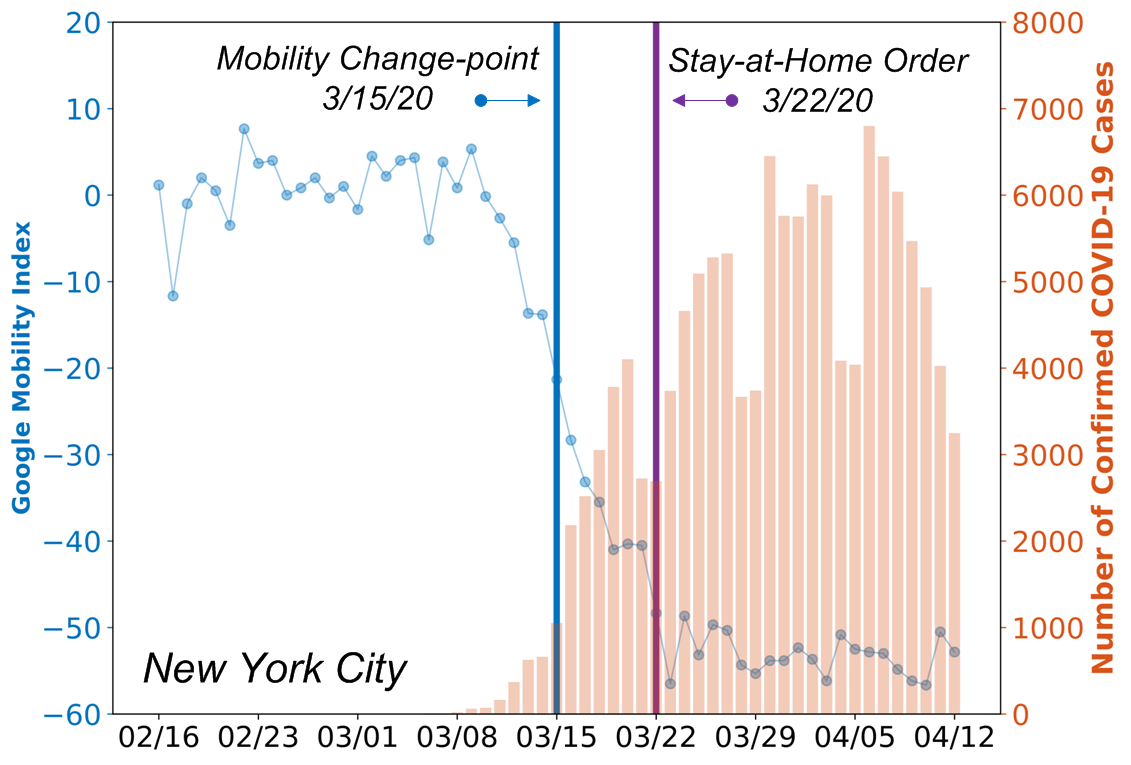}
            \caption[]{}%
            \label{fig:cp_nyc}
        \end{subfigure}
        \caption[ The average and standard deviation of critical parameters ]
        {Detected mobility change-points, Stay-at-Home order dates, daily new COVID-19 cases, and Google Mobility Index in (a) Los Angeles and (b) New York City.} 
        \label{fig:changepoint}
\end{figure*}

In this section, we first show the detected mobility change-points and then discuss our analysis of COVID-19's impact on traffic accidents from three perspectives: inequality, irregularity, and severity.
\subsection{Change-point Detection}
The change-point detection is conducted via solving the optimization function defined in Equation~\ref{eqn:opt}. Fig.~\ref{fig:changepoint} shows the detected mobility change-point dates, the dates when Stay-at-Home orders are issued, the daily Google mobility index, and the number of confirmed COVID-19 cases in Los Angeles and New York City. As a result, for both cities, the detected mobility change-point date is March 15, 2020, which is prior to the announcement of the Stay-at-Home orders. This result indicates people's early response to the pandemic, which coincides with some previous findings for example by Villas-Boas et al.~\cite{villas2020we} but contradicts to other findings where people are observed to adjust their mobility after government announcing guidelines and orders~\cite{ghader2020observed,wang2020spatial}. The inconsistency of the mobility change-point at different regions highlights the necessity of a systematic approach, such as ours, especially consider the change-point is the cornerstone for assessing the impact of the pandemic on various social sectors.

\subsection{Inequality}
The pandemic is likely to affect different demographic groups unequally. In this section, we analyze traffic accidents grouped by various demographic factors, including age, race, gender, etc., before and after the detected mobility change-point, i.e., March 15. Two types of studies are conducted: one associated with traffic accident counts and the other associated with traffic accident fractions. Three study periods are being analyzed, namely 15 days, 30 days, and 60 days before and after the detected mobility change-point in both Los Angeles and New York City. In order to account for possible seasonal shifts in traffic accidents, we also use traffic accident data of both cities during the same time periods in 2019 as references. All analyses reported in this section are conducted via the DID method introduced in Sec.~\ref{sec:methods}.  



\begin{figure*}
        \centering
        \begin{subfigure}[b]{0.5\textwidth}
            \centering
            \includegraphics[width=\textwidth]{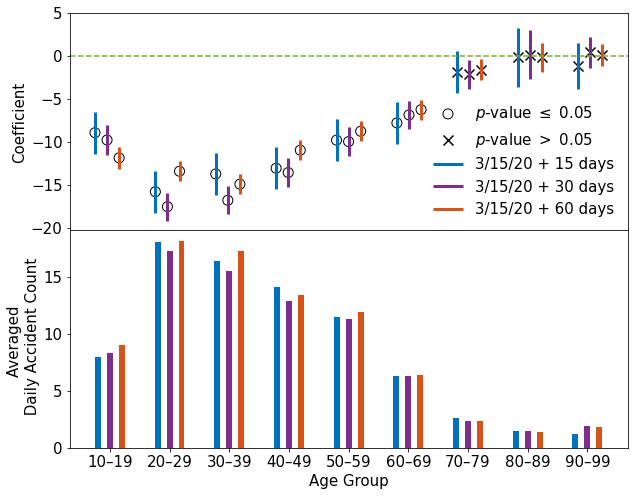}
            \caption[]{}
        \end{subfigure}\hspace*{0.9em}
        \begin{subfigure}[b]{0.5\textwidth}  
            \centering 
            \includegraphics[width=\textwidth]{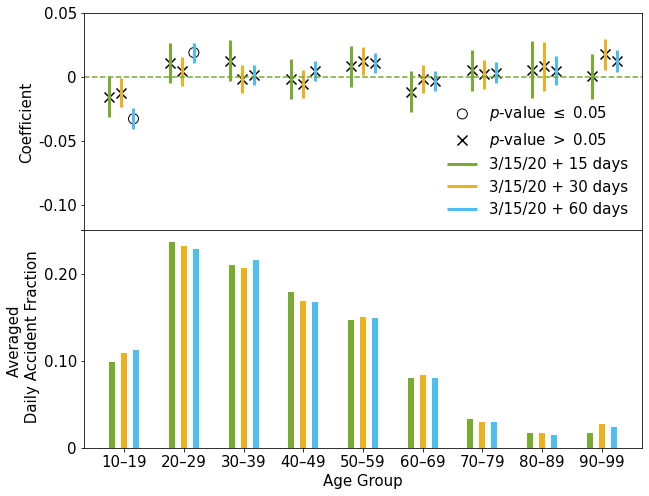}
            \caption[]{}
        \end{subfigure}
        \caption
        {The analysis results of traffic accidents by age. In (a), we can see that all age groups except ``70--79'', ``80--89'', and ``90--99'' have significant reductions in accident counts. The groups ``20-29'' and ``30-39'' have the largest reductions compared to the other groups. In (b), we can observe that the changes of fractions are more balanced. Only the groups ``10--19'', ``20--29'', and ``90--99'' in certain studying periods show significant changes. Lastly, the youngest group ``10--19'' has seen the largest reduction.   } 
        \label{fig:age}
\end{figure*}

Fig.~\ref{fig:age}(a) shows the analysis results regarding different age groups. All age groups except ``70--79'', ``80--89'', and ``90--99'' have significant negative coefficients, indicating reductions of daily traffic accidents in these groups. For seniors older than 70, the coefficient changes are insignificant, showing virtually no reduction of traffic accidents among them. This may due to that seniors in general travel less, hence are involved in fewer accidents regardless of the reduction in overall traffic. Examining the coefficient magnitudes, the age groups ``20-29'' and ``30-39'' have relatively larger reductions compared to the other age groups. 



The analysis results regarding accident fractions are shown in Fig.~\ref{fig:age}(b). The changes of the coefficients are much milder compared to the analysis results regrading accident counts. Only the age groups ``10--19'', ``20--29'', and ``90--99'' in the 60-day analysis (for age group ``90--99'', also the 30-day analysis) show significant changes in fractions. This result indicates that the ratio of traffic accidents between different age groups largely remained the same under the pandemic. Nevertheless, the daily fraction of traffic accidents in the ``10--19'' group is the largest and reduced by 3.2\%.

\begin{figure}
		\centering
		\includegraphics[width=\linewidth]{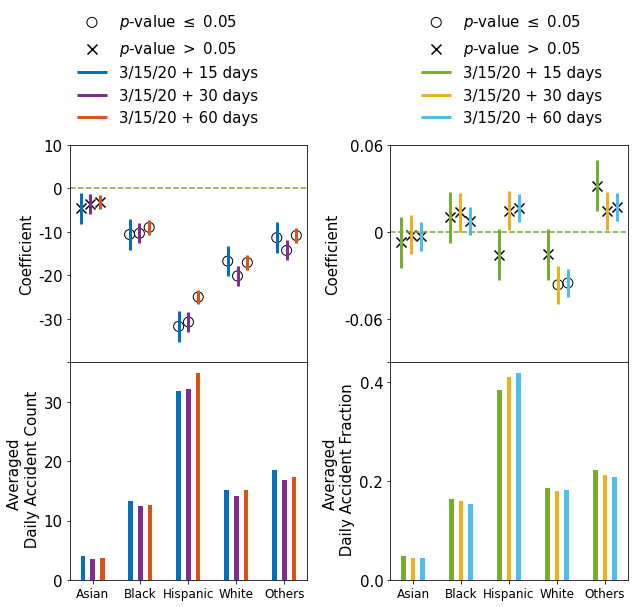}
		\caption{The analysis results of traffic accidents by race. First, there exist reductions in accident counts for all groups. The ``Asian'' group has the lowest reduction in counts and the lowest averaged daily traffic counts. In comparison, the ``Hispanic'' group has the largest reduction in counts and the largest averaged daily counts. In terms of the fractions, the ``White'' group has a reduction in both the 30-day and 60-day analysis, while the ``Black'', ``Hispanic'', and ``Others'' have an increase in nearly all studying periods.}
		\label{fig:race}
\end{figure}

Fig.~\ref{fig:race} shows the analysis results of traffic accidents grouped by race. First, all groups, except ``Asian'', experience significant reduction after the mobility change-point. Second, the ``Asian'' group has the lowest reduction of the coefficients and also the lowest averaged daily traffic accident counts. In comparison, the ``Hispanic'' group has the largest reduction in terms of the coefficients and also the largest averaged daily traffic accident counts. From the fraction analysis, the ``White'' group has a reduction in both the 30-day and 60-day analysis, while the ``Black'', ``Hispanic'', and ``Others'' have an increase in nearly all studying periods.



\begin{figure}
		\centering
		\includegraphics[width=\linewidth]{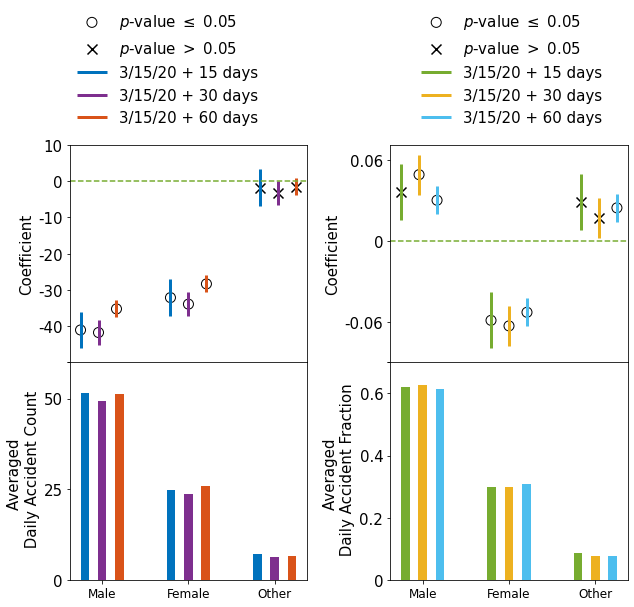}
		\caption{The analysis results of daily traffic accidents by gender. In terms of the counts, both ``Male'' and ``Female'' groups have significant reductions. However, in terms of the fractions, the ``Male'' and ``Other'' groups have increased around 4\%, while the ``Female'' group has decreased around 5\%. }
		\label{fig:gender}
\end{figure}

The analysis results of daily traffic accidents grouped by gender are shown in Fig~\ref{fig:gender}. Regarding the accident counts, both ``Male'' and ``Female'' groups have significant reductions. However, in terms of the fractions, the ``Male'' group has increased around 4\%, while the ``Female'' group has decreased around 5\%. This highlights the unbalanced ratio between different gender groups in traffic accidents under the pandemic. 


\subsection{Irregularity}

\begin{figure*}
        \centering
        \begin{subfigure}[b]{0.5\textwidth}
            \centering
            \includegraphics[width=\textwidth]{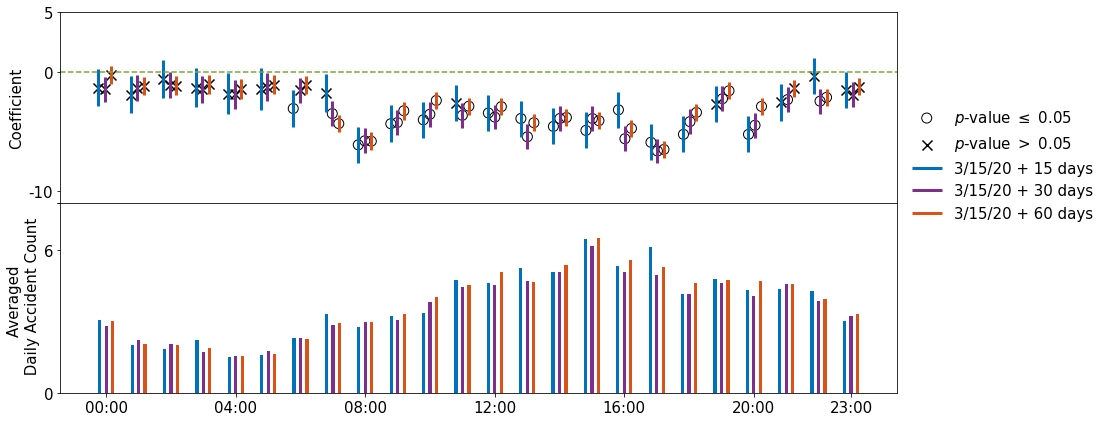}
            \caption[]{}
        \end{subfigure}\hspace*{0.9em}
        \begin{subfigure}[b]{0.5\textwidth}  
            \centering 
            \includegraphics[width=\textwidth]{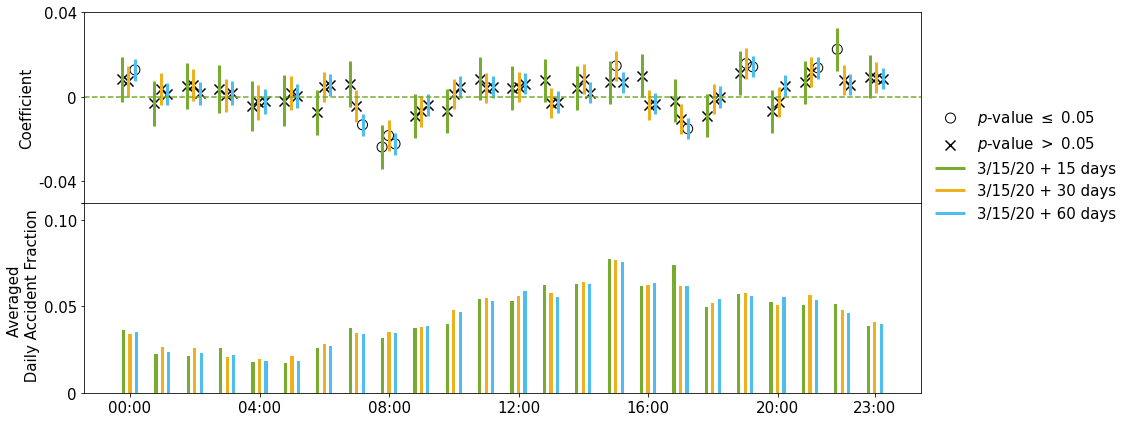}
            \caption[]{}
        \end{subfigure}
        \caption
        {The analysis results of daily traffic accidents grouped by different hours of the day in Los Angeles. In (a), we can see that the accident count has reduced over most daytime and evening hours between 06:00 and 22:00. In (b), we can observe that the traffic accident ``peak'' hours have shifted from usually morning and afternoon hours to late-evening hours. These results demonstrate the time irregularity of traffic accidents during COVID-19.} 
        \label{fig:hours}
\end{figure*}

Fig.~\ref{fig:hours}(a) shows the analysis results of daily traffic accident counts grouped by different hours of the day. For the most part, the hours between 06:00 and 22:00 have seen significant decreases in terms of averaged daily accident account. Fig.~\ref{fig:hours}(b) shows the analysis results of daily traffic accident fractions grouped by different hours of the day. We can see that the fractions are reduced during the regular morning and afternoon peak hours such as 08:00, 17:00, and 18:00; while the fractions after 18:00 are increased. These irregularities showing mobility patterns persistently deviate from commonly observed accident patterns over the daytime hours and peak hours prior to the pandemic.      



In order to test any irregularity of traffic accidents in space, we conduct Kernel Density Estimation (KDE) over a 30-day period before and after the detected mobility change-point, in Los Angeles and New York City, respectively. The 2019 data are also used to account for the seasonal shift of traffic accident patterns. 

\begin{figure*}
		\centering
		\includegraphics[width=\textwidth]{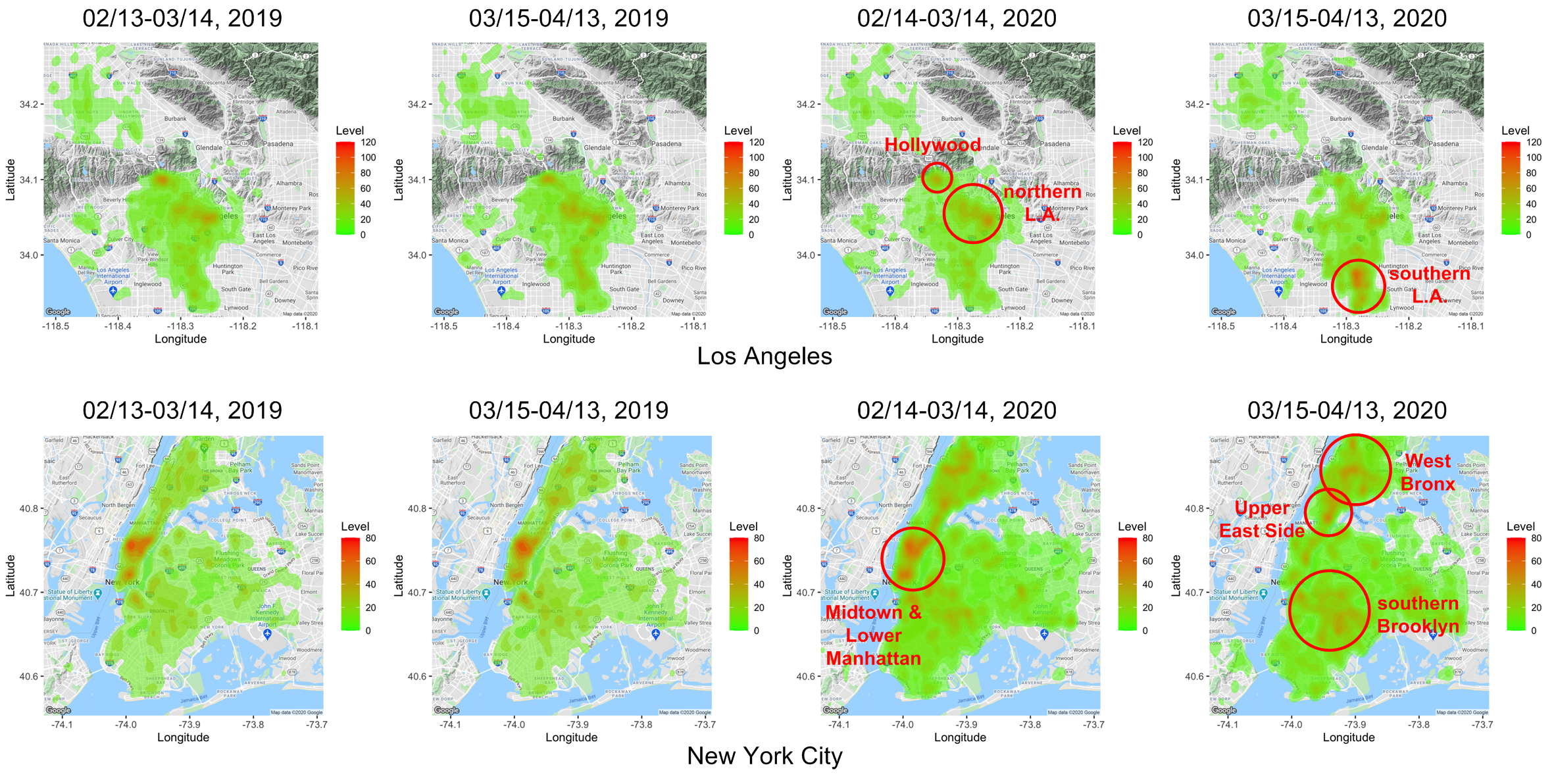}
		\caption{The KDE results of traffic accidents in Los Angeles (TOP) and New York City (BOTTOM). Four 30-day analyses are shown before and after the mobility change-point, i.e., March 15, in 2019 and 2020, respectively. In Los Angeles (L.A.), the accident hotspots have shifted from the Hollywood area and northern L.A. to southern L.A. In New York City, the accident hotspots have shifted from Midtown and Lower Manhattan to Upper East Side, West Bronx, and southern Brooklyn. These results have demonstrated the spatial irregularity of traffic accidents under the pandemic.}
		\label{fig:kde-all}
\end{figure*}

The analysis results of Los Angeles are shown in Fig.~\ref{fig:kde-all} TOP. There are two main traffic accident hotspots in Los Angeles prior to the pandemic: one around the Hollywood area and the other around northern Downtown Los Angeles. The total monthly traffic accidents are as high as 80 cases. Some less apparent hotspots are scattered in southern Los Angeles. In contrast, under the pandemic, the most prominent hotspot has shifted from northern Los Angeles to southern Los Angeles, with the number of traffic accidents increased to 110.

The results of New York City are shown in Fig.~\ref{fig:kde-all} BOTTOM. Prior to the pandemic, the accident hotspots appear at Midtown Manhattan and Lower Manhattan. Due to the impact of the COVID-19, Upper East Side, West Bronx, and southern Brooklyn now show more traffic accidents.

\begin{figure*}
        \centering
        \begin{subfigure}[b]{0.45\textwidth}
            \centering
            \includegraphics[width=\textwidth]{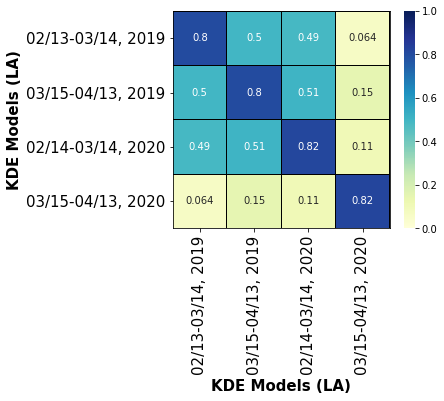}
            \caption[]{}%
            \label{fig:kde_pvalue_la}
        \end{subfigure}\hspace*{0.9em}
        \begin{subfigure}[b]{0.45\textwidth}  
            \centering 
            \includegraphics[width=\textwidth]{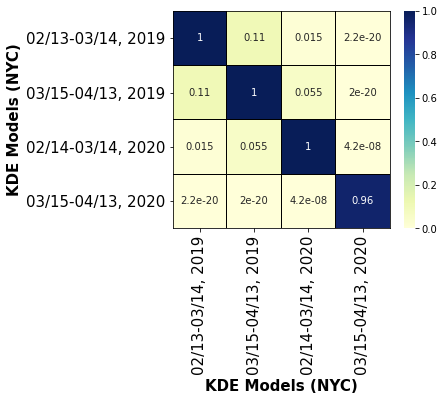}
            \caption[]{}%
            \label{fig:kde_pvalue_nyc}
        \end{subfigure}
        \caption[ The average and standard deviation of critical parameters ]
        {The global two-sample comparison test~\cite{duong2012closed} of our KDE results in two cities. The $p$-values after the detected mobility change-point are lower than the $p$-values of other time periods, indicating the irregularity of traffic accident patterns during COVID-19.} 
        \label{fig:kde-pvalue}
\end{figure*}

In addition to the visualization of the KDEs, we conduct the global two-sample comparison test~\cite{duong2012closed} defined in Equation~\ref{eqn:ISE} to quantitatively compare the KDE results. Fig.~\ref{fig:kde-pvalue} shows the $p$-values of our statistical tests in both cities.
The $p$-values between the time period after the detected mobility change-point and all other time periods are significantly lower than the rest of the $p$-values, which indicates the corresponding KDE distributions are  different and confirms our qualitative visual inspection results.  

\subsection{Severity}

\begin{figure*}
        \centering
        \begin{subfigure}[b]{0.33\textwidth}
            \centering
            \includegraphics[width=\textwidth]{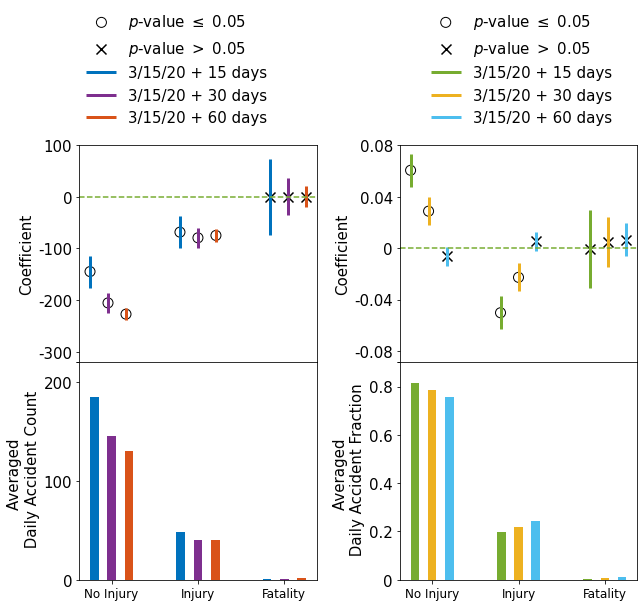}
            \caption[]{}
        \end{subfigure}\hspace*{0.9em}
        \begin{subfigure}[b]{0.33\textwidth}  
            \centering 
            \includegraphics[width=\textwidth]{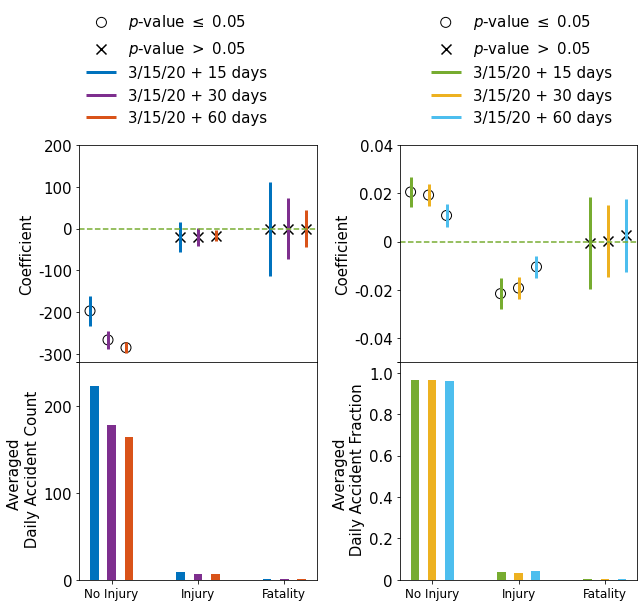}
            \caption[]{}
        \end{subfigure}\hspace*{0.9em}
        \begin{subfigure}[b]{0.33\textwidth}  
            \centering 
            \includegraphics[width=\textwidth]{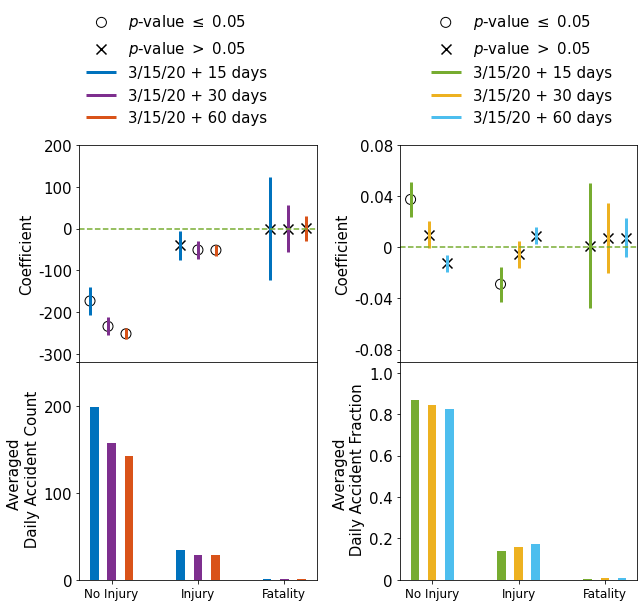}
            \caption[]{}
        \end{subfigure}
        \caption
        {The analysis results of accident severity in New York City: (a) accidents that do not involve other transportation modes; (b) accidents that involve pedestrians; and (c) accidents that involve motorists. For all three cases, the number of no-injury and injury cases are decreased. However, the number of fatality cases remains the same. The fraction analysis shows that the no-injury cases are relatively higher than the injury cases, until 60 days after the mobility change-point.} 
        \label{fig:injury}
\end{figure*}

In this section, we report our analysis results of the pandemic's impact on accident severity. First, Fig.~\ref{fig:injury}(a) shows that the coefficients of ``Fatality'' are insignificant over all three study periods, which implies no change of the fatality rate in traffic accidents even the total number of accidents has decreased. In addition, both the ``No Injury'' and ``Injury'' categories have seen a reduction, while the former decreases more than the latter. Similar patterns are observed in Fig.~\ref{fig:injury}(b) and Fig.~\ref{fig:injury}(c): the ``No Injury'' category experiences larger decreases compared to the ``Injury'' category, and no change of the  ``Fatality'' category is observed. The fraction analysis shows that the no-injury cases in the 15-day and 30-day analyses are higher than the injury cases. Such a difference disappears in the 60-day analysis.


\section{Conclusion}
The ongoing pandemic has significantly impacted every aspect of our society. In this study, we analyze the influence of COVID-19 on road traffic safety through the example cities Los Angeles and New York City. Specifically, we have analyzed traffic accidents associated with different demographic groups, different times of the day, spatial distribution, and severity levels of accidents that involve or do not involve pedestrians and motorists. We have found that 1) the pandemic has disproportionately affected certain age groups, races, and genders; 2) the accident ``hotspots'' have been shifted in both time and space compared to prior-pandemic time periods; and 3) the number of non-fatal accident cases has seen a reduction, however the number of fatal accident cases remains the same under the pandemic. 

We are mainly interested in exploring two research directions in the future. The first direction is to be able to simulate traffic accidents under the pandemic. In this way, we can test safety intervention policies effectively and efficiently. To facilitate this research direction, we can leverage existing data processing techniques~\cite{Li2017CityEstSparse,Lin2019ComSense}, estimation and prediction methods~\cite{Li2018CityEstIter,lin2018predicting}, and simulation algorithms~\cite{Wilkie2015Virtual,Li2017CityFlowRecon}. The other research direction is to study when traffic flows are undergoing a drastic change, how can shared mobility and/or autonomous vehicles be used to reduce traffic accidents. The analyses conducted in this study can provide some insights on how to distribute and allocate new transportation modalities. The proposed strategies can be evaluated and tested via recent advances in autonomous driving modeling and simulation~\cite{Li2019ADAPS,Chao2019Survey}.

\FloatBarrier




%
{\small
	\bibliographystyle{IEEEtran}
	\bibliography{reference}
}

%



\end{document}